\documentstyle[12pt]{article}

\def\lromn#1{\uppercase\expandafter{\romannumeral#1}}

\def\blist{\begin{list}{\setlength{\rightmargin}{\leftmargin}}}
\def\elist{\end{list}}
\begin{document}
\begin{flushright}
%\today \\
TU/99/577\\
RCNS-99-05\\
%hep-ph/9910***
\end{flushright}

\begin{center}
\begin{large}

\bf{
Temperature Power Law of Equilibrium Heavy Particle Density
}

\end{large}

\vspace{36pt}

\begin{large}
Sh. Matsumoto and M. Yoshimura

Department of Physics, Tohoku University\\
Sendai 980-8578 Japan\\
\end{large}

\vspace{4cm}

{\bf ABSTRACT}

\end{center}

A standard calculation of the energy density of
heavy stable particles that may pair-annihilate into
light particles making up thermal medium
is performed to second order of coupling, using the technique of
thermal field theory. At very low temperatures a power law of
temperature is derived for the energy density of the heavy particle.
This is in sharp contrast to the exponentially suppressed contribution 
estimated from the ideal gas distribution function.
The result supports a previous dynamical calculation based on
the Hartree approximation, and implies that the
relic abundance of dark matter particles is enhanced
compared to that based on the Boltzmann equation.

%%%%%%%%%%%%%%%%%%%%%%%%%%%%%%%%%%%%%%%%%%%%%%%%%%%%%

\newpage

\vspace{0.5cm} 
\hspace*{0.5cm} 
Estimate of the relic number density of dark matter
particles such as LSP (Lightest Supersymmetric Particle) is
conventionally made using a thermally averaged Boltzmann
equation \cite{lee-weinberg}. 
When the freeze-out temperature $T_{f}$ is relatively high, 
for instance $T_{f} > M/5$
with $M$ the mass of the annihilating heavy particle,
this procedure is justified, but only after a careful examination
(as we shall do in this paper).
On the other hand, if the temperature is low, e.g, for
$T_{f} < M/30$ (a case typical for the cold dark matter), 
the use of the Boltzmann equation is dubious,
and a more general quantum kinetic equation becomes necessary
\cite{my-pair-99}.
A basic reason for this is that a finite time behavior of
quantum system should be analyzed beyond the Boltzmann
equation which  uses the on-shell S-matrix element, hence
is not fundamental at the full quantum mechanical level.
As is well known, 
quantum mechanics at finite time cannot be described
in terms of the S-matrix alone.

Some model calculation using a kinetic equation
based on the Hartree approximation was performed in \cite{my-pair-99},
assuming slow variation of the occupation number.
In cosmology the temperature variation is given by
the adiabatic law, $\dot{T}/T = -\,$ the Hubble
rate, and it was found that the heavy particle number 
density follows the equilibrium value until a freeze-out 
temperature below which the annihilation is frozen.
This picture of the sudden freeze-out \cite{lee-weinberg} is valid both
for the Boltzmann equation and for our new kinetic equation.
Thus, estimate of the freeze-out temperature using
the equilibrium abundance is
crucial even when one includes off-shell effects.
It has been shown \cite{my-pair-99}, \cite{jmy-decay} that 
for the equilibrium occupation number 
higher order term in coupling dominates  at low temperatures 
over that given by the ideal gas distribution function
($1/(e^{E/T} - 1)$ for bosons).
A temperature power dependence was thus derived
for the equilibrium heavy particle abundance.
A larger relic abundance than previously thought of emerged, hence
a more restrictive region for the model parameter space is
anticipated.

In the present work we employ a more familiar technique
of the thermal field theory (imaginary-time formalism).
Although there is no real time in the thermal field theory that
governs the out-of equilibrium dynamics,
the thermal average one computes here deals with manifestly off-shell
quantities. Hence the present method departs from the S-matrix
approach of the thermally averaged Boltzmann equation.
(In fairness we should point out that our kinetic approach
\cite{my-pair-99} justifies the thermally averaged Boltzmann
equation at high temperatures, but not at low temperatures.)
We shall derive for the observable energy density 
a basically similar, but numerically different equilibrium result, 
from the previous result \cite{my-pair-99};
the temperature power is different.
Since definition of the occupation number is somewhat
ambiguous in interacting field theory, a direct computation
of the energy density is desirable,
which we do in the present work.
In a companion paper \cite{my-pair-99-3}
we derive a new form of the kinetic
equation, the one much simpler in form than that of ref.\cite{my-pair-99}.
The idea there is the Hartree approximation using the influence
functional method, and we shall clarify how the detailed
balance equation there agrees with the result given in the
present work.

For simplicity, we take throughout this paper a relativistic
boson model, assuming the annihilation interaction
\begin{equation}
\frac{\lambda }{4}\,\varphi ^{2}\,\chi ^{2} \,, 
\end{equation}
with $\varphi $ being the heavy particle field and 
$\chi $ being the light field taken as making up a part of
the thermal bath.
The themal average of any operator $O$ is done by
using the Gibbs weight $e^{-\,\beta H}$ with
$\beta = 1/T$ the inverse temperature,
\begin{equation}
\langle O \rangle \equiv \frac{{\rm tr}\: O\,e^{-\,\beta H}}
{{\rm tr}\; e^{-\,\beta H}} - \left( T=0 \;{\rm contribution}\right)
\,.
\end{equation}
The total hamiltonian $H$ contains both contribution from
the thermal bath and interaction.
Both for consistent renormalization and for thermalization
of light $\chi $ particles, we need self-interaction of the
form,
\begin{equation}
\frac{\lambda _{\varphi }}{4!}\,\varphi ^{4} \,, 
\hspace{0.5cm} 
\frac{\lambda _{\chi }}{4!}\,\chi ^{4} \,.
\end{equation}
We numerically assume $|\lambda _{\varphi }|$ to be much less than
$\lambda ^{2}$, 
but $\lambda _{\chi }$ of arbitrary order provided
$|\lambda | \ll |\lambda _{\chi }| < 1$.
This hierarchy of coupling constants is assumed keeping in mind that
the annihilation interaction is weak, while lighter particles
can be kept in thermal equilibrium due to a stronger self-interaction
given by $\lambda _{\chi}$.
The mass hierarchy $m_{\chi } \ll m_{\varphi }$ also helps in favor of
this picture.

We compute the $\varphi $ energy density $\rho _{\varphi }$
including renormalization counter terms;
\begin{eqnarray}
&&
\rho _{\varphi } = \langle {\cal H}_{\varphi } 
+ ({\rm counter \; terms}) \rangle 
\,, 
\\ &&
{\cal H}_{\varphi } = -\,\frac{1}{2}\, \dot{\varphi }^{2} +
\frac{1}{2}\, (\nabla \varphi )^{2}  + \frac{1}{2}\, M^{2}\,
\varphi ^{2}  \,,
\end{eqnarray}
where $\dot{\varphi }$ is the Euclidean time derivative.
Note that by our definition of the thermal trace the zero-temperature
value is automatically subtracted in this formula.

To leading order $O[\lambda ^{0}]$ the $\varphi $ energy density
given by this definition contains the "free field" result;
\begin{equation}
\rho _{\varphi }^{(0)} = \int\,\frac{d^{3}p}{(2\pi )^{3}}\,
\frac{\sqrt{p^{2} + M^{2}}}{e^{\sqrt{p^{2} + M^{2}}/T} - 1} \,.
\end{equation}

The quantity $\rho _{\varphi }$ contains product of 
field operators at the same spacetime point
and require renormalization at the composite
operator level, separately from the Lagrangian counter terms. 
Complication due to the operator mixing also occurs.
For instance, the operator mixing of composite operators occurs among
nine second rank tensor operators of mass dimension up to 4,
\begin{equation}
g_{\mu \nu }\,\left( \,
\varphi ^{2} \,, \chi ^{2} \,, 
\varphi ^{2}\chi ^{2} \,, \varphi ^{4} \,, \chi ^{4}
\,, (\partial \varphi)^{2} \,, (\partial \chi )^{2} 
\,\right) \,, \hspace{0.2cm}
\partial _{\mu }\varphi \partial _{\nu }\varphi \,, \hspace{0.2cm}
\partial _{\mu }\chi \partial _{\nu }\chi 
\,.
\end{equation}
The renormalized Hamiltonian ${\cal H}_{\varphi }$ is then
of the form,
\begin{equation}
{\cal H}_{\varphi } = {\cal H}_{\varphi }^{{\rm bare}} +
\sum_{i}^{2}\,Z_{i}^{(2)}\,O_{2\,, i} + 
\sum_{i}^{7}\,Z_{i}^{(4)}\,O_{4\,, i}
\end{equation}
to $O[\lambda ^{2}]$.
There are two operators $O_{2\,, i}$ of mass dimension 2, while
there are seven operators $O_{4\,, i}$ of mass dimension 4.
There is no $ Z^{(1)}$ factor of order $\lambda $ due to
the assumed interaction hierarchy.

Renormalization constants $Z_{i}$ are derived by assigning a
canonical value to the zero-temperature
Fourier transformed correlation function at one external momentum point.
(We use the on-shell point for renormalization.)
For example, for 
$ \partial _{\mu }\varphi \partial _{\nu }\varphi  $
the correlators with field operators up to mass dimension 4
\begin{eqnarray}
&&
\langle 0| \partial _{\mu }\varphi \partial_{\nu } \varphi 
(x)\,\psi (x_{1})\psi (x_{2})|0 \rangle
 \,,
\\ &&
\langle 0| \partial _{\mu }\varphi \partial_{\nu } \varphi 
(x)\,\psi (x_{1})\psi (x_{2})
\psi (x_{3}) \psi (x_{4})|0 \rangle
 \,,
\end{eqnarray}
with $\psi $ either $\varphi $ or $\chi $, should be taken into account.
(Higher dimensional correlators are all finite and irrelevant
to renormalization.)
The free field limit values are used to define $Z_{i}$.

To order $O[\lambda ^{2}]$ there are three types of diagrams
(two topologically distinct)
contributing to the energy density $\langle {\cal H}_{\varphi } \rangle$. 
See Fig.1 and Fig.2.
The temperature power dependence arises from the amputated sunset
diagram (Fig.1), which gives in the configuration space 
\begin{eqnarray}
\lambda ^{2}\,\int_{0}^{\beta }\,d^{4}x\,d^{4}y\,
\,\Delta _{\varphi }(x)\Delta _{\varphi }(-y)\Delta _{\varphi }(y-x)
\Delta _{\chi  }(x-y)\Delta _{\chi  }(y-x)
\,, 
\label{sunset contribution0} 
\end{eqnarray}
prior to renormalization.
The propagator in thermal medium
$\Delta_{\phi\,, \,\chi }(y)$ is periodic in the Euclidean time $y_{0}$ with
a period $\beta = 1/T$, hence giving the range of integration 
$0 \sim  \beta $ as explicitly indicated. 
The Fourier transformed propagator has the well known form,
$\Delta (\omega _{n} \,, \vec{p}) \sim 1/(-\,\omega _{n}^{2} + \vec{p}^{2}
+ M^{2})$ with discrete $\omega _{n} = 2\pi i\,n/ \beta $
($n = 0 \,, \pm 1 \,, \pm 2 \,, \cdots $).
The other contributions from Fig.2 are exponentially
suppressed by $e^{-\,M/T}$ with $M$ the heavy $\varphi $ particle mass,
the factor familiar in the conventional approach.

As usual, one rewrites eq.(\ref{sunset contribution0}) using
the Fourier transform.
The resulting discrete energy sum over $\omega _{n} $
can be converted to a contour integral of this variable $z = \omega _{n}$,
using the function $1/(e^{\beta z} - 1)$.
After some algebraic manipulation, one finds that to $O[\lambda ^{2}]$
\begin{eqnarray}
\rho _{\varphi }^{(2)}  &\sim &
  -\,\lambda^2\,\int dk~dk'~dp~dp'
  (2\pi)^3\delta(\vec{p} + \vec{p}' + \vec{k} + \vec{k}')\,2\omega _{p}
  \nonumber \\
  \nonumber \\
  &~&
  \left[\,
    \frac{f_pf_{p'}(1 + f_k)(1 + f_{k'})
      -(1 + f_p)(1 + f_{p'})f_kf_k'}
    {(\omega_p + \omega_{p'} - \omega_k - \omega_{k'})^2}
  \right.
  \nonumber \\
  \nonumber \\
  &~&
  +\,2\,
  \frac{f_pf_{p'}(1 + f_k)f_{k'}
    -(1 + f_p)(1 + f_{p'})f_k(1 + f_k')}
  {(\omega_p + \omega_{p'} - \omega_k + \omega_{k'})^2}
  \nonumber \\
  \nonumber \\
  &~&
+\,
  \frac{f_pf_{p'}f_kf_{k'}
    -(1 + f_p)(1 + f_{p'})(1 + f_k)(1 + f_k')}
  {(\omega_p + \omega_{p'} + \omega_k + \omega_{k'})^2}
  \nonumber \\
  \nonumber \\
  &~&
+\,2\,
  \frac{f_p(1 + f_{p'})(1 + f_k)(1 + f_k')
    -(1 + f_p)f_{p'}f_k f_{k'}}
  {(\omega_p - \omega_{p'} - \omega_k - \omega_{k'})^2}
  \nonumber \\
  \nonumber \\
  &~&
\left.
+\,2\,
    \frac{f_p(1 + f_{p'})f_k(1 + f_{k'})
      -(1 + f_p)f_{p'}(1 + f_k)f_k'}
    {(\omega_p - \omega_{p'} + \omega_k - \omega_{k'})^2}
    \,  \right]
 \,. \label{sunset contribution} 
\end{eqnarray}
We dropped minor Boltzmann suppressed terms to obtain this result.
A shorthand notation for the phase space integral
$dk = d^3k/(2\pi)^32\omega_k$
was used here, and $f_{p\,,\, p'}$ are the occupation number for
the heavy $\varphi $ particle, while $f_{k\,,\, k'}$ are that for
the light $\chi $ particle;
\begin{equation}
f_{p} = \frac{1}{e^{\sqrt{p^{2} + M^{2}}/T} - 1} \,, \hspace{0.5cm} 
f_{k} = \frac{1}{e^{k/T} - 1} \,.
\end{equation}
A similar form to eq.(\ref{sunset contribution}) was derived for
the proper self-energy in ref.\cite{thermal self-energy}. 
For simplicity we asumme that the $\chi $ mass $m_{\chi } \ll T$, and
indeed take $m_{\chi } = 0$ here.

Terms containing $f_{p}$ or $f_{p'}$ in eq.(\ref{sunset contribution}) 
are Boltzmann suppressed by $e^{-M/T}$.
Dropping all these Boltzmann suppressed terms,
one obtains after removal of the infinity 
\begin{eqnarray}
  \frac{\lambda^2}{16\pi^2}\int dk~dk'~f_k~f_{k'}\,(\omega _{k}^{2}
  + \omega _{k'}^{2})
  \int^\infty_{4M^2} ds\,\frac{1}{s^{2}} 
  \sqrt{1 - \frac{4M^2}{s}}\,
 \,.
\end{eqnarray}
This gives to leading order of $T/M$
\begin{equation}
\rho _{\varphi }^{(2)} = c\,\lambda^2\,\frac{T^6}{M^2} \,, \hspace{0.5cm} 
c =    \frac{1}{69120}  \sim 1.4 \times 10^{-5}
\,.
\end{equation}
Terms of higher temperature-power are subleading, and
neglected here.

There is a simple reason how the temperature dependence
$\propto T^{6}$ arises.
Since ${\cal H}_{\varphi }$ has both dimension 2 and 4 operators,
terms of order $T^{2}$ and $T^{4}$ are divergent and they are
cancelled by counter terms $Z_{i}\,O_{i}$.
The reason why one does not have divergent $O[\lambda ^{2}T^{6}/M^{2}]$
terms is that $Z_{i}$ is already of order $\lambda ^{2}$ and one
only needs $\langle O_{i} \rangle$ to $O[\lambda ^{0}]$
which is either Boltzmann-suppressed or has no $T^{6}M^{-2}$ term.
The remaining finite term is then of order $T^{6}$ unless some special
cancellation mechanism works.
In our companion paper \cite{my-pair-99-3} 
we give a separate computation of the equilibrium
energy density from the kinetic approach, which exactly agrees 
with the present result.

We separately computed the interaction energy density 
given by Fig.3 to get
\begin{equation}
\rho _{{\rm int}} 
= 
\langle \frac{\lambda }{4}\,\varphi ^{2}\chi ^{2} +
({\rm counter \; terms}) \rangle 
\sim -\,\frac{\pi^2}{64800}\,\lambda ^{2}\,\frac{T^{8}}{M^{4}}  \,,
\end{equation}
dropping $O[\lambda]$ Boltzmann suppressed terms.
Thus, the interaction makes a minor contribution in the
pair-annihilation model.

How about terms of order $\lambda \lambda _{\chi }^{n}$ ?
All these turn out to give Boltzmann suppressed contributions
to $\rho _{\varphi }$ and $\rho _{{\rm int}}$.
This makes our result insensitive to the self-interaction among
light $\chi $ particles.

A possibility that the interaction makes at low temperatures 
a comparable contribution to the heavy particle energy 
was noted by Singh and Srednicki \cite{singh-srednicki} who explicitly
calculated the interaction energy 
in the simple solvable model of \cite{jmy-decay}.
However their suggestion that quantum kinetic approach should be
discarded and we should go back to the Boltzmann equation
is not valid for a number of reasons.
First of all, there is no solid justification
for the Boltzmann equation at low temperatures.
Next, the amount of interaction energy is model dependent;
indeed our annihilation model gives a negligible contribution of
interaction. We are neither confident of their claim that
the solvable gaussian model gives a comparable contribution
of interaction, becasuse renormalization is not
taken into account in their computation.
Moreover,
as we shall show below, one can make in the annihilation model 
a clear distinction between various forms of energy by
taking into account the cosmological expansion.
This argument makes clear which part should be regarded as
the dark matter energy density.

We would however like to point out some peculiarity;
cancellation of order $\lambda ^{2}\,T^{6}$ terms for the total energy,
 $\rho _{{\rm tot}} = \rho _{\varphi } + \rho _{\chi } +
\rho _{{\rm int}}$.
Namely, a term in $\rho _{\chi }$ of order $\lambda ^{2}\,T^{6}$
exactly cancells the same order term in $\rho _{\varphi }$ to
give a $\lambda ^{2}\,T^{8}$, and no $\lambda ^{2}\,T^{6}$, 
term to $\rho _{{\rm tot}}$.
This however does not mean that both, $\rho _{\chi }$ and 
$\rho _{{\rm tot}}$, are numerically of order $\lambda ^{2}\,T^{8}$.
There are more important, larger terms of order, for instance,
$\lambda \lambda_{\chi } ^{n}$ for these.
With a larger $\lambda _{\chi }$ coupling, one has a consistent
picture for thermalization of $\chi $ particles.

Although it is technically difficult to compute the momentum
distribution of relic particles, it is relatively easy to
compute the pressure of heavy particles in thermal equilibrium.
It turns out that the pressure $p$ is one fifth of the energy,
$p = \frac{1}{5}\,\rho $. It is neither of completely non-relativistic
(in which case $p = 0$) 
nor of relativistic ($p = \frac{1}{3}\,\rho$) form.
Thus, right after the freeze-out 
the equation of state implies that the energy density follows
$\rho \propto a^{-\,18/5}$.
But the well known redshift effect makes the high momentum component
energetically subdominant after rapid cosmologial expansion.
It is thus reasonable to suppose that at later epochs the heavy
particle is essentially non-relativistic, behaving like
$\rho _{\varphi } \propto a^{- 3}$.
On the other hand, 
the interaction energy density $\approx  \lambda \varphi ^{2}\chi ^{2}$
decreases much more rapidly like $a^{-\,5}$ in thermal equilibrium.
This decrease is faster than the high momentum part of $\rho _{\varphi }$.
We thus find that the correct dark matter density decreasing with
the volume factor should be identifed as $\rho _{\varphi }$.

Since our formula for the energy density is valid only at low temperatures,
$T \ll M$, meaning that the indivisual particle energy
$E \sim M$, one can also deduce for the heavy particle
number density at $T \ll M$ 
\begin{equation}
n_{\varphi } = \frac{\rho _{\varphi }^{(0)} + \rho _{\varphi }^{(2)}}{M} 
\sim  \frac{\rho _{\varphi }^{(0)} }{M} 
+ c\,\lambda ^{2}\,\frac{T^{6}}{M^{3}} \,.
\label{eq number density} 
\end{equation}

There is a critical temperature $T_{cr}$ below which the temperature
power term dominates over the usual Boltzmann term.
This may easily be estimated by equating the two formulas;
\begin{equation}
(\frac{MT_{cr}}{2\pi })^{3/2}\,e^{-M/T_{cr}} =
c\lambda ^{2}\,\frac{T_{cr}^{6}}{M^{3}} \,.
\end{equation}
Numerically, the value of $T_{cr}/M$ ranges from $1/28$ to $1/33$ for
$\lambda = 0.1 - 0.01$.
Thus, in this $\lambda $ range $T_{cr} \approx M/30$, very crudely.
A useful empirical formula of the critical temperature
in the coupling range of $\lambda = 0.1 - 10^{-4}$ is
\begin{equation}
\frac{M}{T_{{\rm cr}}} = 23 - 2.3\,\ln \lambda \,.
\end{equation}

Is our result reliable at the zero temperature limit, $T \rightarrow 0$ ?
We argue that this is not so from the following reason.
Our method of using the equilibrium value for the freeze-out abundance
is based on available sufficient time for relaxation towards equilibrium.
In cosmology expansion makes this time limited to the Hubble time.
The physical time scale towards equilibrium is
the inverse of the pair-creation rate for 
$\chi \chi \rightarrow \varphi \varphi $,
\begin{equation}
\langle \sigma v \rangle\,n_{\varphi } \approx 
O[10^{-7}]\,\lambda ^{4}\,(\frac{T}{M})^{5}\,T \,.
\end{equation}
This should be compared to the Hubble rate, 
$O[1]\,N^{1/2}\,T^{2}/m_{{\rm pl}}$ where $N$ is the relativistic
degrees of freedom contributing to the energy density.
The condition for relaxation in cosmology is then
\begin{equation}
\frac{T}{M} \gg O[10^{-3}]\,\frac{N^{1/8}}{\lambda }(\frac{M}{100\,
GeV})^{1/4}
\,.
\end{equation}
For a very small $\lambda $ this condition may violate
$T < M$, in which case
the use of the ordinary Boltzmann equation is justified.
We thus need a relatively large coupling and/or a relatively small
mass $M$ for our new result to be dominant over the Boltzmann
suppressed number density.
In our published calculation of \cite{my-pair-99} 
there is a technical mistake in computation of
the $\varphi $ number density so that
the correct condition for the off-shell effect becomes more stringent.
The numerical result there should thus be corrected, and 
the parameter region for the new effect is much more
reduced, as seen more fully in our subsequent analysis.

We now estimate the freeze-out temperature $T_{f}$.
The simplest way is to equate the Hubble rate given as
a function of the temperature $T$ to the annihilation rate
$\langle \sigma\,v  \rangle\,n_{\varphi }$, where $\sigma $ is
the annihilation cross section of order $\lambda ^{2}/M^{2}$.
Using the number density, eq.(\ref{eq number density}) 
and equating to the Hubble rate
\begin{equation}
H = 1.66\times \sqrt{N}\,\frac{T^{2}}{m_{{\rm pl}}} \,, 
\end{equation}
we find for $T_{f} < T_{cr}$
\begin{equation}
T_{f} \approx 0.3\, GeV\,\frac{N^{1/8}}{\lambda }\,(\frac{M}{100\,
GeV})^{5/4}
\,.
\end{equation}
This holds only when the temperature power term dominates over
the exponential $e^{-M/T}$ term for $n_{\varphi }$.
The relic mass density is then given by
\begin{equation}
\left( \frac{n_{\varphi }}{T^{3}} \right)_{f} \approx 
4\times 10^{-13}\,\frac{N^{3/8}}{\lambda}\,
(\frac{M}{100\,GeV})^{3/4}
\,.
\end{equation}
Demanding that this is smaller than the present critical density 
gives a constraint on the model parameter, 
the coupling $\lambda $ and the heavy mass $M$.
The $\varphi $ mass density relative to the critical density
is at present
\begin{equation}
\frac{\rho _{\varphi }^{(2)}}{\rho _{c}} \approx 10^{-4}\,
\frac{N^{3/8}}{\lambda}\,(\frac{M}{100\,GeV})^{7/4} \,.
\end{equation}
Numerical estimate of the freeze-out temperature including
the Boltzmann suppressed region is given in Fig.4.
The relativistic degree of freedom $N$ is taken 10.75 throughout
this paper.
At a given $\lambda $ this freeze-out temperature substantially differs
from the naive estimate obtained by using the Boltzmann factor,
when the heavy particle mass is small.

A more precise, yet approximate estimate is possible by using
time evolution equation in the expanding universe.
From the equation for the number density
\begin{equation}
\frac{dn_{\varphi }}{dt} + 3\,H \,n_{\varphi } =
-\,\langle \sigma v \rangle\,(\,n_{\varphi }^{2} - n_{{\rm eq}}^{2}\,)
\,, 
\label{time evol for n} 
\end{equation}
one has for the yield $Y = n_{\varphi }/T^{3}$
\begin{eqnarray}
&&
\frac{dY}{dT} = d\,\langle \sigma v \rangle\,m_{{\rm pl}}\,
(\,Y^{2} - Y_{{\rm eq}}^{2}\,) \,, 
\label{time evolution eq} 
\\ &&
d = 1.66\,\sqrt{N} \,, \hspace{0.5cm} 
\langle \sigma v \rangle \approx \frac{\lambda ^{2}}{32\pi \,M^{2}}
 \,,
\end{eqnarray}
where the temperature-time relation $t \propto T^{-2}$ was used.
The equilibrium value is approximately 
a sum of two terms valid at high (but $T/M \ll 1/\sqrt{c}\lambda 
\approx 3\times 10^{2}/\lambda $) and low temperatures,
\begin{equation}
Y_{{\rm eq}} = \frac{1}{T^{3}}\,\int\,\frac{d^{3}p}{(2\pi )^{3}}\,
\frac{1}{e^{\sqrt{p^{2} + M^{2}}/T} - 1} + c\,
\lambda ^{2}\,(\frac{T}{M})^{3} \,.
\end{equation}
In Fig.5 we plot an example of numerical computation of the time
evolution equation (\ref{time evolution eq}).
Among two examples of coupling $\lambda = 0.05 \,, 0.3$
both with $M = 100\, GeV$, the smaller coupling case gives
indistiguishable new effect and the ordinary Boltzmann approach
is numerically correct.
On the other hand, the larger coupling case gives a substantially
different relic abundance from the conventional result.

Time evolution obtained from eq.(\ref{time evol for n}) supports
the picture of sudden freeze-out as in the Lee-Weinberg analysis
\cite{lee-weinberg}.
The major difference here is the new equilibrium abundance $n_{{\rm eq}}$.

The final relic abundance of dark matter particles is shown
as a contour plot in the parameter ($M \,, \lambda $) plane.
In Fig.6 we show the present $\varphi $ mass density relative
to the critical density, for computation both with and without
our new effect.
Our new effect tends to show up for a larger coupling and a smaller
mass.

Since our new effect gives an additional positive contribution to 
an energy integral,
the relic density is always enhanced from the conventional
result without our effect.
Thus the allowed parameter region in the model parameter space
gets always smaller by our result.
The real question is then how much of the previously known region
is excluded by this new effect.
How our result of the relic density affects the presumably best
motivated case of the SUSY dark matter ramains to be studied numerically.
Especially in the $Z$ and the Higgs resonance region the effective
coupling is large \cite{resonance effect in relic abundance}, 
and there the off-shell effect may  be very
large. This should be checked in more realistic calculation
beyond our boson model.

\vspace{1cm}
%\newpage
\begin{center}
{\bf Acknowledgment}
\end{center}

We thank Drs. Singh and Srednicki for communicating
their result prior to publication.
The work of Sh. Matsumoto is partially
supported by the Japan Society of the Promotion of Science.

\newpage
\begin{Large}
\begin{center}
{\bf Figure caption}
\end{center}
\end{Large}

\vspace{0.5cm} 
\hspace*{-0.5cm}
{\bf Fig.1}

Amputated sunset diagram for the composite operator.
The solid line and the broken line are the heavy $\varphi $ particle
and the light $\chi $ particle propagators in thermal equilibrium.
The crossed circle represents the insertion of the composite operator.

\vspace{0.5cm} 
\hspace*{-0.5cm}
{\bf Fig.2}

The rest of diagrams contributing to the composite
operator as in Fig.1 of $O[\lambda ^{2}]$.

\vspace{0.5cm} 
\hspace*{-0.5cm}
{\bf Fig.3}

Diagrams contributing to the interaction energy.

\vspace{0.5cm} 
\hspace*{-0.5cm}
{\bf Fig.4}

The inverse freeze-out temperature $T_{f}^{-1}$ vs the mass $M$
of the heavy particle, computed for two cases of coupling,
$\lambda = 0.1 \,, 0.05$.
Dotted curves are result using the Boltzmann factor.

\vspace{0.5cm} 
\hspace*{-0.5cm}
{\bf Fig.5}

Time evolution of the relative energy density of heavy particles,
$M\,n_{\varphi }/n_{\gamma }$,
computed for two cases of coupling, $\lambda = 0.3 \,, 0.05$.
For the larger coupling ($\lambda = 0.3$) our new result and
old result based on the thermally averaged Boltzmann equation
do substantially differ, while the smaller coupling case
gives indistinguishable result.

\vspace{0.5cm} 
\hspace*{-0.5cm}
{\bf Fig.6}

Contour plot of the present mass density of relic particles.
The lines shown correspond to 
$1 \,, 10^{-1} \,, 10^{-2} \times $ the critical
mass density $\rho _{c}^{0}$ (taken here $1.05 \times 10^{-5}\,
h_{0}^{2}\,GeV\,cm^{-3}$ with $h_{0} = 0.7$), 
along with results obtained using the Boltzmann suppression factor.

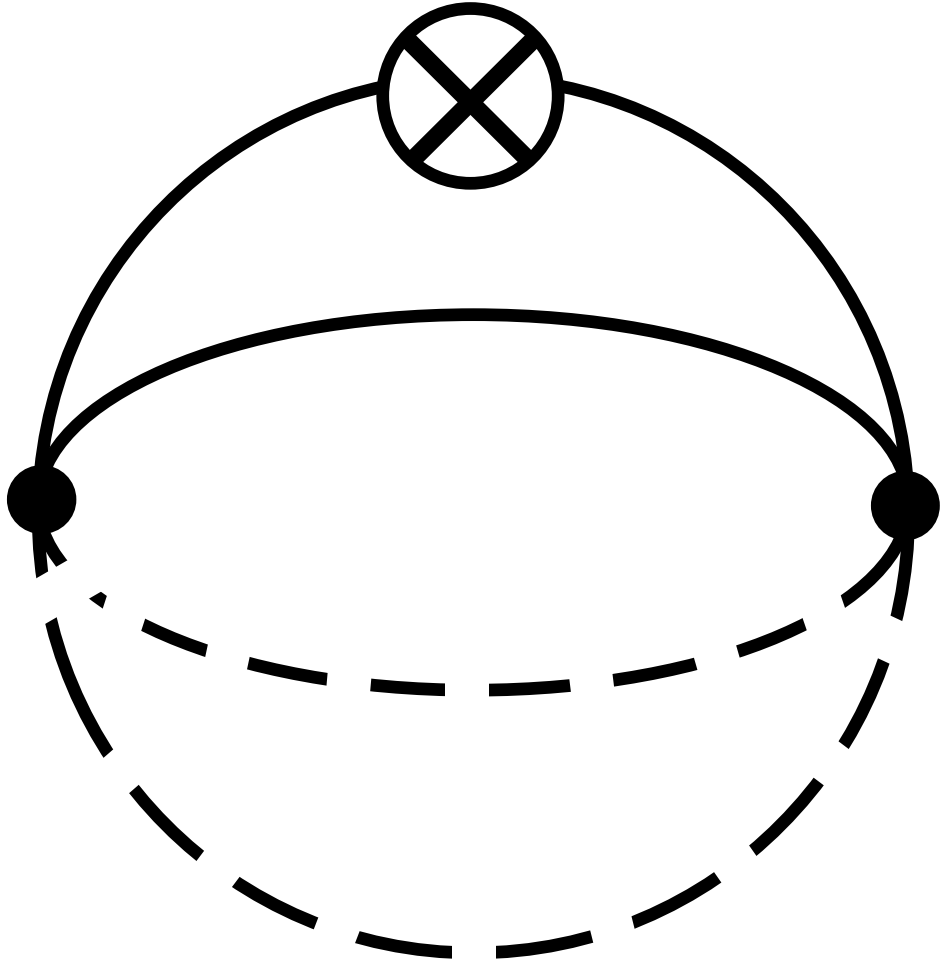
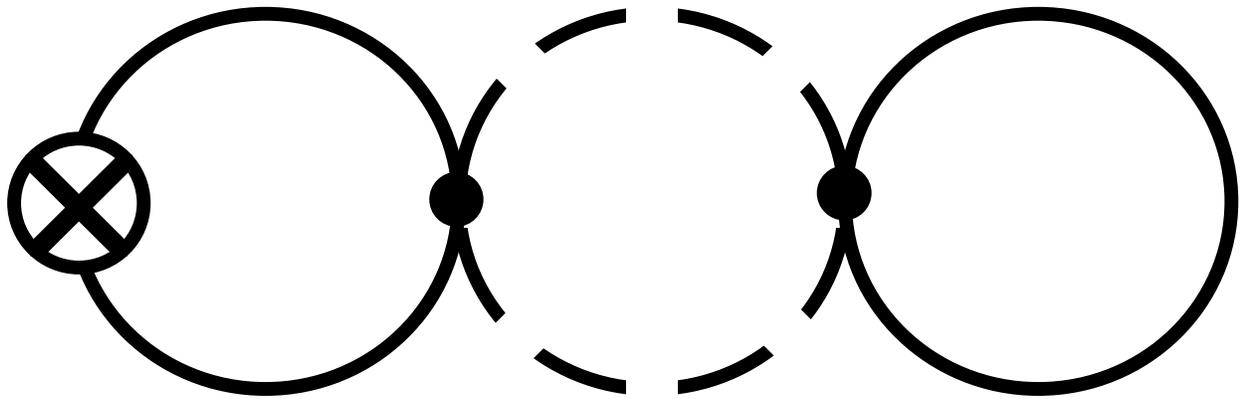
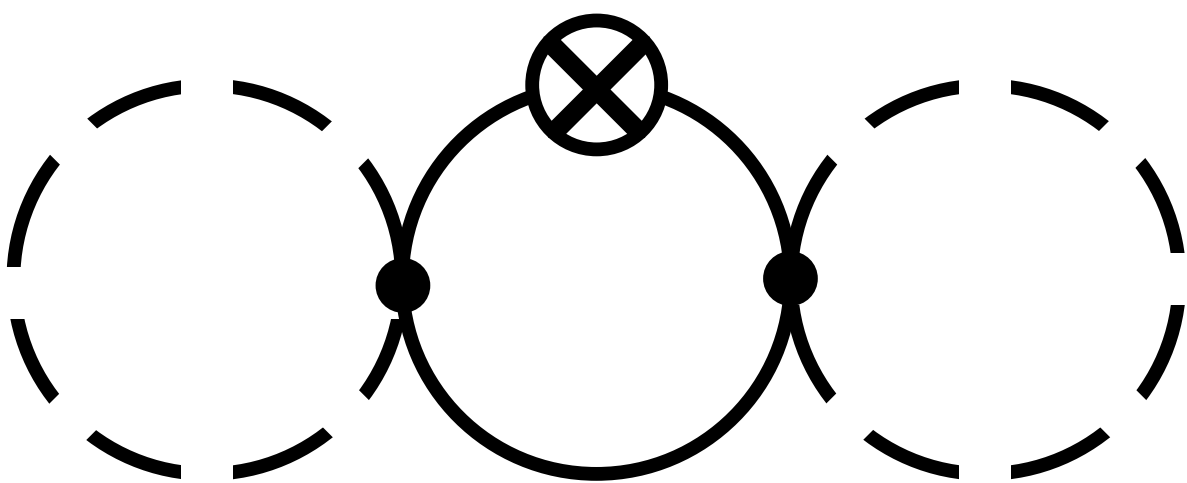
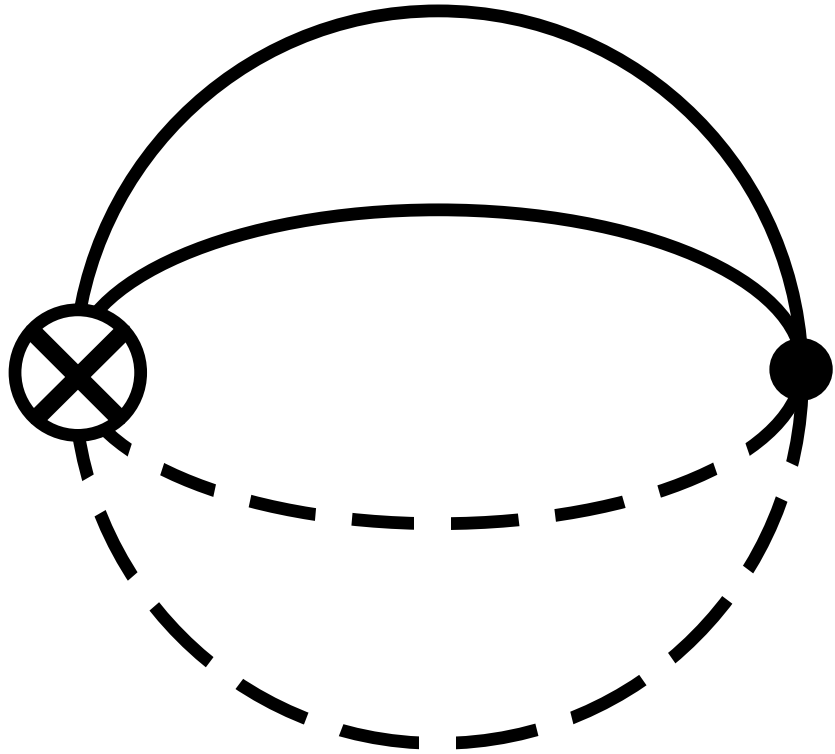
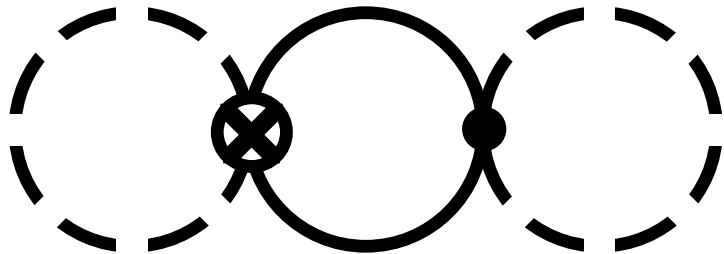
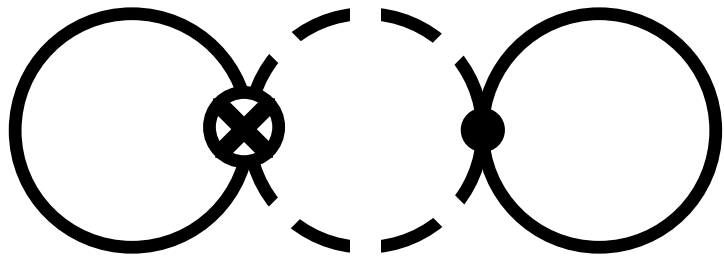
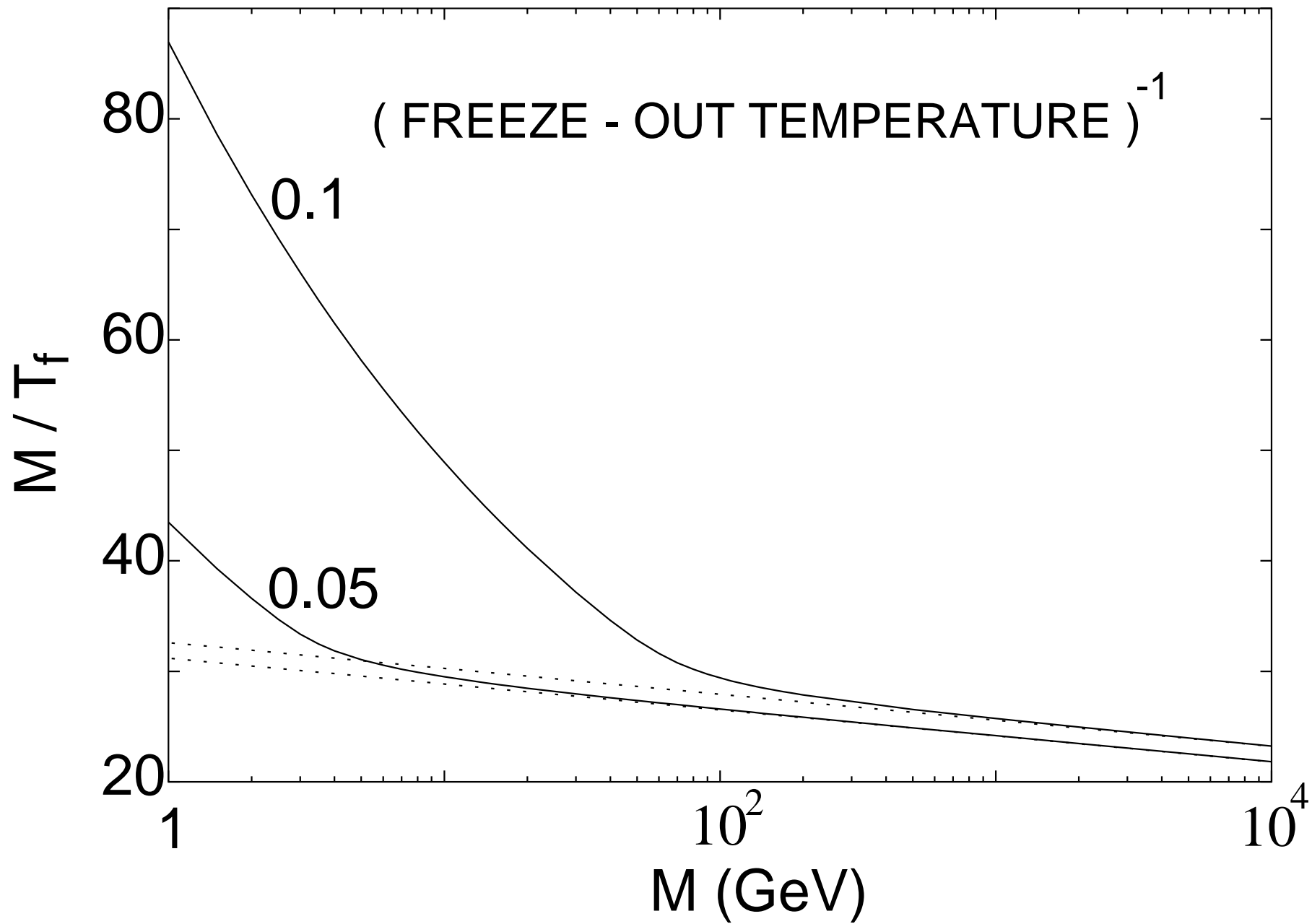
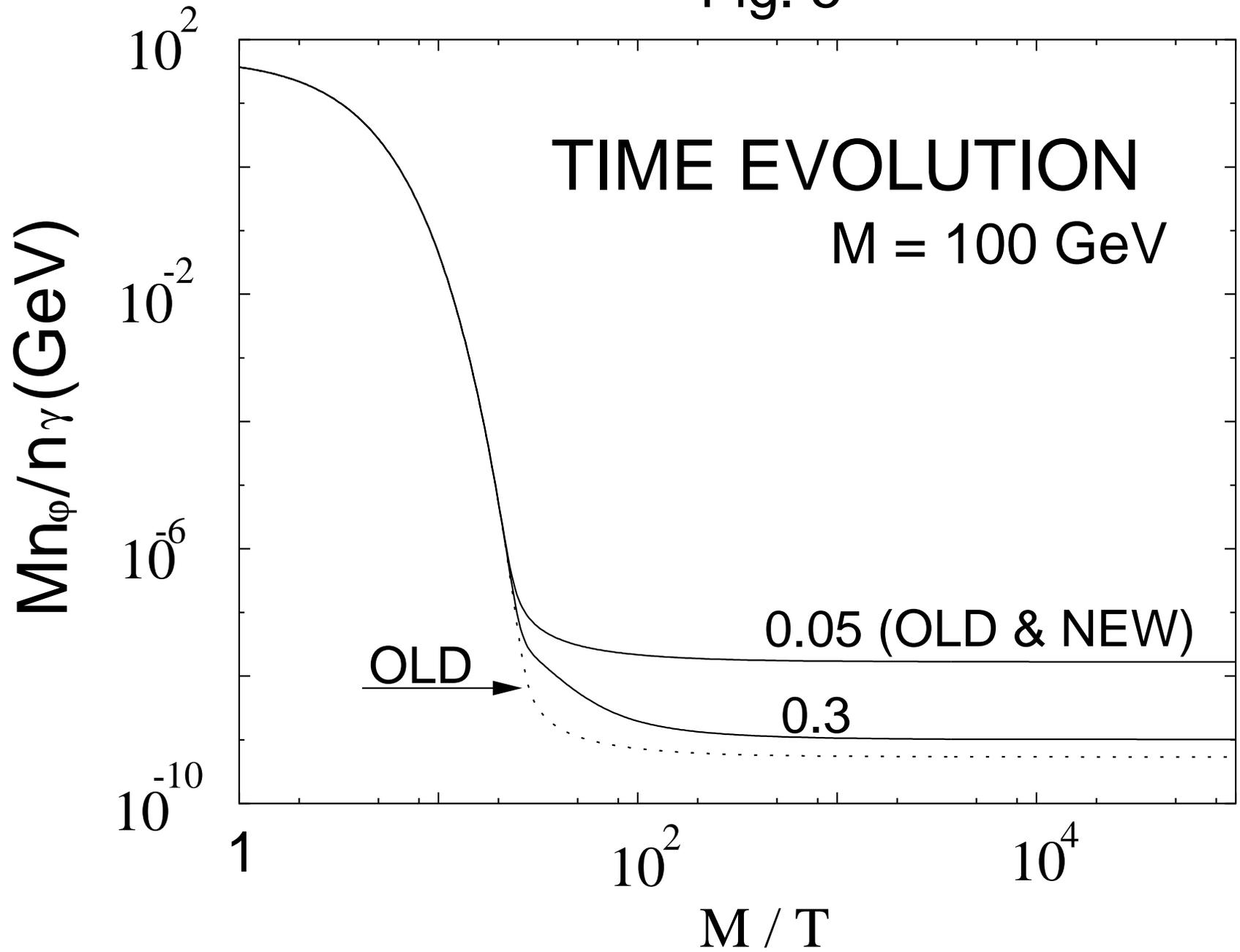
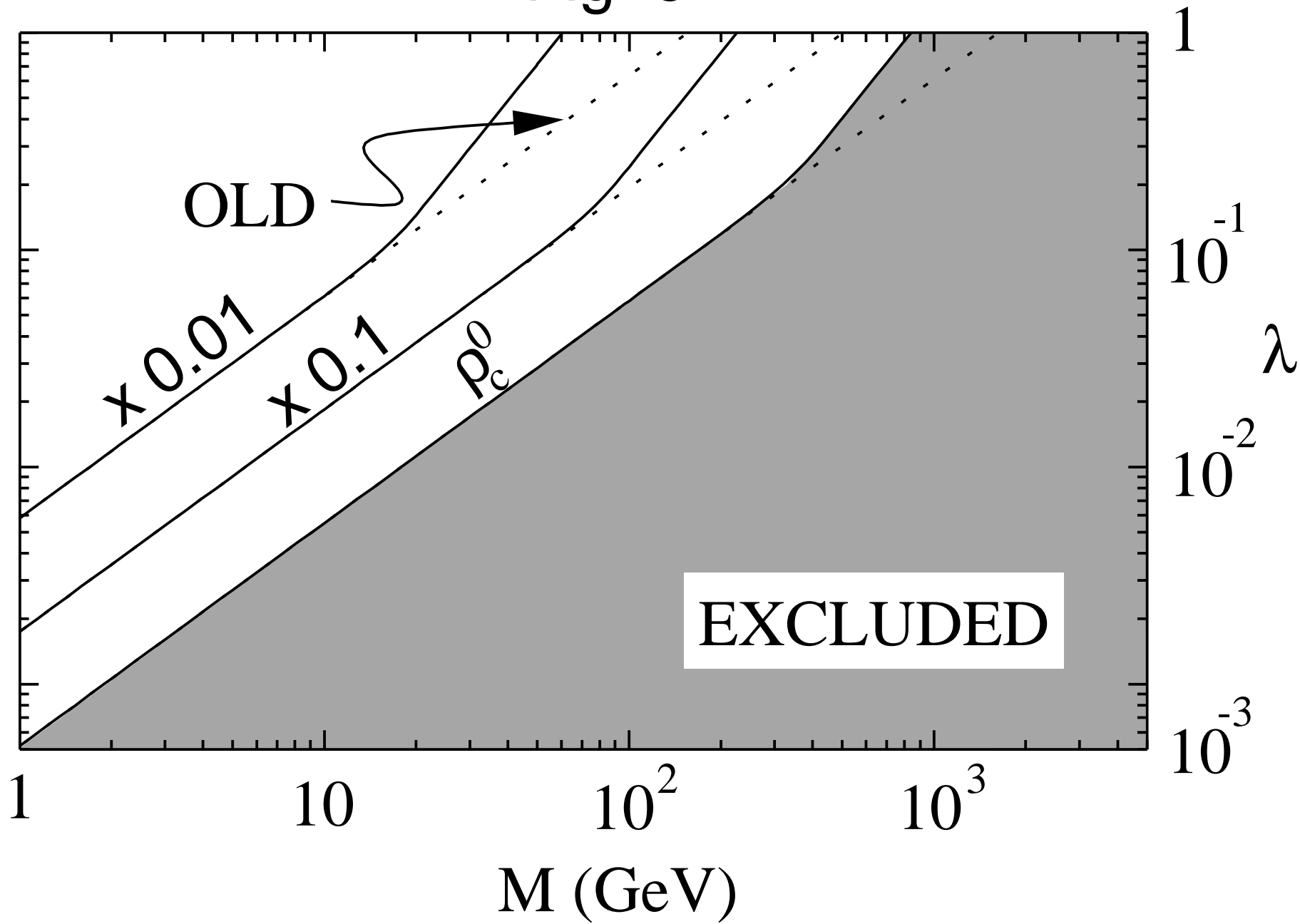
\end{document}